\documentstyle[11pt,a4]{article}

\newcommand{\eqalign}[1]{
\null \,\vcenter {\openup \jot \ialign {\strut \hfil $\displaystyle {
##}$&$\displaystyle {{}##}$\hfil \crcr #1\crcr }}\,}
\newcommand{\be}{\begin{equation}}
\newcommand{\ee}{\end{equation}}
\newcommand{\ba}{\begin{array}}
\newcommand{\ea}{\end{array}}
\newcommand{\bea}{\begin{eqnarray}}
\newcommand{\eea}{\end{eqnarray}}
\newcommand{\pa}{\partial}

\newcommand{\AmS}{{\protect\the\textfont2
  A\kern-.1667em\lower.5
ex\hbox{M}\kern-.125emS}}

\title{\begin{flushright}{\normalsize
hep-th/9606187}
\end{flushright}
\vspace{2 cm} 
Quantum corrections of Abelian
        Duality Transformations}

\author{J. Balog\thanks{E-mail: balog@mppmu.mpg.de}\\ 
       {\small MPI,
        M\"unchen, 
        P.O. Box 40 12 12, M\"unchen, Germany}\\
        P. Forg\'acs\thanks{E-mail: forgacs@celfi.phys.univ-tours.fr}\\ 
        {\small Laboratoire de Math.~et Physique~Th\'eorique,
        Universit\'e de Tours},\\
        {\small Parc de Grandmont, F-37200 Tours,
        France}\\
        Z. Horv\'ath\thanks{E-mail: zalanh@ludens.elte.hu}
       and L.Palla\thanks{E-mail: palla@ludens.elte.hu}\\
        {\small Institute for Theoretical Physics,
        Roland E\"otv\"os University,} \\
       {\small  H-1088, Budapest, Puskin u. 5-7, Hungary}
}

\begin{document}
\maketitle
\begin{abstract}
A modification of the Abelian Duality transformations is proposed
guaranteeing that a (not necessarily conformally invariant) $\sigma$-model
be quantum equivalent (at least up to two loops in perturbation theory)
to its dual. This requires  
a somewhat non standard perturbative treatment of the {\sl dual}
$\sigma$-model.
Explicit formulae of the modified duality transformation are presented for a
special class of block diagonal purely metric $\sigma$-models.
\end{abstract}
\vspace{11 mm}
\begin{center}
key words: sigma models, duality, quantum corrections
\end{center}
\vfill\eject
\section{Introduction}
Various T (`target space') duality transformations \cite{busch},
\cite{review}, \cite{phrep} connecting two seemingly
different $\sigma$-models
or string-backgrounds have aroused a considerable amount of interest.
In Ref.~\cite{bfhp} 
we have investigated quite a few examples of dually related 
`ordinary' (i.e.\ not necessary conformally invariant) 
$\sigma$-models treated as 
two dimensional quantum field theories in the framework of 
perturbation theory.  We have shown on a 
number of examples that the `naive' (tree level) T-duality
transformations \cite{busch} {\sl cannot be}
exact symmetries of the quantum theory. The `naive' Abelian duality
transformations yield a model equivalent to the original one only 
to one loop order in perturbation theory, however, the equivalence
breaks down in general,
at the two loop order. We reached these conclusions by 
comparing various $\beta$ functions in the original and dual theories.
Therefore it seems to be clear
that the question of quantum equivalence between dual 
$\sigma$-models deserves further study.
 
The quantum equivalence of dually related 
Conformal Field Theories (CFT) has been proven 
in Ref.~\cite{rove}. For  a number of special cases
(Wess-Zumino-Witten (WZW) models \cite{kiri},
gauged WZW models \cite{kiri2} and two dimensional black holes \cite{tse0})
this equivalence has been shown explicitly. 
 
To fix ideas let us first write
the two dimensional $\sigma$-model action:
\be\eqalign{
\label{gsm}
S=\frac{1}{4\pi\alpha'}\int d^2z
\biggl[&\sqrt{h}h^{\mu\nu}\left(g_{00}\pa_\mu\theta \pa_\nu\theta
+2g_{0\alpha}\pa_\mu\theta\partial_\nu \xi^{\alpha} 
+g_{\alpha\beta}\partial_\mu
\xi^{\alpha}\partial_\nu \xi^{\beta}\right)\cr
&+i\epsilon^{\mu\nu}(2b_{0\alpha}\pa_\mu\theta \partial_\nu \xi^{\alpha}
+b_{\alpha\beta}\partial_\mu
\xi^{\alpha}\partial_\nu \xi^{\beta})\biggr]
}\ee
where $ g_{ij}$ is the target space metric, 
$b_{ij}$ is the (antisymmetric) torsion potential,
$h_{\mu\nu}$ is the world sheet metric and
$\alpha^{'}$ the inverse of the string tension.
In Eq.~(\ref{gsm}) we have assumed that there is a Killing vector 
and in the adopted coordinate system,
the target space indices are decomposed as $i=(0,\alpha)$
corresponding to splitting the coordinates as $\xi^i=(\theta, \xi^\alpha)$,
and then the background fields $(g\,,b)$ are independent of the coordinate
$\xi^0=\theta$. Note the absence of the dilaton field in Eq.~(1).
In this letter we concentrate mainly on not conformally invariant     
 $\sigma$-models, quantized as ordinary quantum field  theories. 
In the same spirit 
the world sheet metric, $h_{\mu\nu}$, is taken to be
flat in what follows. 

Now the well known formulae of Abelian T-duality
\cite{busch}, mapping the `original' $\sigma$-model with action,
$S[g\,,b]$, given in Eq.~(\ref{gsm}) to its dual, $S[\tilde g\,,\tilde b]$
 are: 
\be\eqalign{
\label{dmetr}
&{\tilde g}_{00}={1\over g_{00}}\qquad
{\tilde g}_{0\alpha}={b_{0\alpha} \over g_{00}}\,,
\qquad
{\tilde b}_{0\alpha}={g_{0\alpha} \over g_{00}}\,, \cr
&{\tilde g}_{\alpha\beta}=g_{\alpha\beta} -
{g_{0\alpha}g_{0\beta} - b_{0\alpha} b_{0\beta}\over g_{00}}\,,\quad
{\tilde b}_{\alpha\beta}=b_{\alpha\beta}-{g_{0\alpha}b_{0\beta}
         -g_{0\beta}b_{0\alpha}\over g_{00}}\,.
}\ee
It has been recently found that the Abelian
duality transformation rules (\ref{dmetr})
can be recovered in an elegant way -- without ever using the dilaton -- by
performing a canonical transformation \cite{alglo}. This clearly shows
that the models
related by these transformations are  {\sl classically} equivalent. In the 
quantum theory, the usual way to argue that the dually
related models are equivalent  
in spite of the non linear change of variables involved,
is by making some formal manipulations in the 
functional integral \cite{busch}, ignoring the need for regularization. 
For a special class of {\sl conformally invariant} $\sigma$-models
(string backgrounds)
it has already been found in Ref.~\cite{tse1} that
the Abelian T-duality transformations rules of Ref.~\cite{busch}
should be modified at the two loop level to preserve conformal invariance. 

The aim of this letter is to put forward a nontrivial modification
of the standard Abelian T-duality transformations, Eqs.~(\ref{dmetr}),
which should promote them to a full {\sl quantum} symmetry.
The basic motivation for such a modification is easy to understand;
the {\sl bare} and the {\sl renormalized} quantities do not transform in the 
same way under duality transformations beyond one loop order in 
perturbation theory. 
While our proposed modification of the T-duality transformation rules
is certainly necessary to ensure
that this symmetry hold in the quantum theory,
it implies 
that the \lq naive' duality 
transformations receive perturbative corrections order by order 
(beyond one loop). Even more interestingly the modified duality
transformations do not map $\sigma$-models into $\sigma$-models
in the usual sense, except for the class of {\sl conformally invariant}
models (or string backgrounds).

We illustrate how the proposed modifications ensure two loop
equivalence between the original and its dual
on an example of an asymptotically free    
$\sigma$-model (the O(3) model) and on the example of two free fields,
written in polar coordinates, both cases treated 
as ordinary quantum field theories.

\section{Modified duality transformations}

When deriving the Abelian duality transformations all formal manipulations
are carried out on  {\sl unrenormalized}, i.e.\ {\sl bare} quantities. 
The partition function of a generic $\sigma$-model using dimensional
regularization can be written as:
\be\label{fint}
Z=
\int D\xi^i\exp\Bigl(-{\mu^{-\epsilon}\over4\pi\alpha^\prime}
\int d^{2-\epsilon}z\,T_{ij}^{(0)}(g,b)\Xi^{ij}\Bigr)\,,
\ee
where $\Xi^{ij}=(\pa_\mu\xi^i\pa^\mu\xi^j
+i\epsilon_{\mu\nu}\pa^\mu\xi^i\pa^\nu\xi^j)$,
and the generalized bare metric, $T_{ij}^{(0)}=g_{ij}^{(0)}+b_{ij}^{(0)}$,
has been computed in terms of the renormalized quantities ($g_{ij}$,
$b_{ij}$) by several authors \cite{hulto}, \cite{metse}, \cite{osb1},
by the background field method in the dimensional regularization
scheme:
\be\label{fint1}
 T_{ij}^{(0)}(g,b)=g_{ij}+b_{ij}+{\alpha^\prime\over\epsilon}
\hat{R}_{ij}(g,b)+
 {(\alpha^\prime)^2\over\epsilon}\hat{Y}_{ij}(g,b)+...\,,
\ee
where 
\be\label{y}
\eqalign{
\hat{Y}_{ij}=&{1\over8}Y^{lmk}_{{\phantom{lmk}j}}\hat{R}_{iklm}\,,\cr
Y_{lmkj}=&-2\hat{R}_{lmkj}+3\hat{R}_{[klm]j}
+2(H^2_{kl}g_{mj}-H^2_{km}g_{lj})\,,\cr
H^2_{ij}=&H_{ikl}H_j^{kl}\,,\qquad
2H_{ijk}=\partial_ib_{jk}+{\rm cyclic}\,.
}\ee 
In Eqs.~(\ref{fint1}, \ref{y}) $\hat{R}_{ij}$
resp.\ $\hat{R}_{iklm}$ denote
the `generalized' Ricci resp.\ Riemann tensors of the
`generalized' connection, $G^i_{jk}$, containing also the torsion term
in addition to the Christoffel
symbols, $\Gamma^i_{jk}$ of the metric $g_{ij}$;
$
G^i_{jk}=\Gamma^i_{jk}+H^i_{jk}\,.
$
A very natural idea would be to perform the  
`naive' duality transformations (\ref{dmetr})
on the bare quantities, $T_{ij}^{(0)}(g,b)$, that is to impose
as the quantum duality symmetry:
\be\label{ndual}
 \tilde{T}_{ij}^{(0)}(g,b)=T_{ij}^{(0)}(\tilde{g},\tilde{b})\,,
 \ee
where \  $\tilde{\ }$ \  denotes the transformation defined by
Eq.~(\ref{dmetr}),
( the symmetric
part of $T_{ij}^{(0)}$, $T_{(ij)}^{(0)}$, transforms as the metric, $g_{ij}$,
while the antisymmetric part, $T_{[ ij] }^{(0)}$, as $b_{ij}$).
For example 
\be
\tilde{T}_{00}^{(0)}(g,b)=\tilde g_{00}^{(0)}={1\over g_{00}
+\alpha^\prime/\epsilon\hat{R}_{00}(g,b)+\dots }=
T_{00}^{(0)}(\tilde g,\tilde b)\,.
\ee
In fact Eqs.~(\ref{ndual}) should only hold modulo diffeomorphisms
(redefinitions of the target space coordinates).

For all the examples studied in \cite{bfhp} it has been  
found that in the one loop order the original and the dual models are 
equivalent after the field redefinition (reparametrization): 
$\xi^i_0\rightarrow\xi^i+\alpha^\prime\xi^i_1(\xi)/\epsilon$,  
$\xi^i_1(\xi)\sim\pa_i\ln g_{00}(\xi)$; (see Eq.~(\ref{renormxi}) below), 
implying that (at least for
the cases in question) Eqs.~(\ref{ndual}) hold.
Comparing the coefficients of ${\alpha^\prime/\epsilon}$ 
on the two sides of (\ref{ndual}) one finds 
that the generalized Ricci tensors computed from
the original and dual quantities should be related -- 
up to a reparametrization -- as:
\be\label{ricci1}
\eqalign{ 
\hat{R}^{\tilde{g}}_{00}&=-{1\over g_{00}^2}
\hat{R}^g_{00},\quad
\hat{R}^{\tilde{g}}_{(0\alpha)}=-{1\over g_{00}^2}\left(
b_{0\alpha}\hat{R}^g_{00}-\hat{R}^b_{[0\alpha] }g_{00} \right)\,,\cr
\hat{R}^{\tilde{b}}_{[ 0\alpha] }&=-{1\over g_{00}^2}\left(
g_{0\alpha}\hat{R}^g_{00}-\hat{R}^g_{(0\alpha)}g_{00} \right)\,,\cr
\hat{R}^{\tilde{g}}_{(\alpha\beta )}&=\hat{R}^g_{(\alpha\beta )}-
{1\over g_{00}}\left( \hat{R}^g_{(0\alpha )}g_{0\beta}+
\hat{R}^g_{(0\beta )}g_{0\alpha}-\hat{R}^b_{[ 0\alpha] }b_{0\beta}-
\hat{R}^b_{[ 0\beta] }b_{0\alpha}\right) \cr
 &+ {1\over
g_{00}^2}\left( g_{0\alpha}g_{0\beta}-b_{0\alpha}b_{0\beta}\right) 
\hat{R}^g_{00}\,,\cr
\hat{R}^{\tilde{b}}_{[\alpha\beta] }&=\hat{R}^b_{[\alpha\beta] }-
{1\over g_{00}}\left( \hat{R}^g_{(0\alpha )}b_{0\beta}+
\hat{R}^b_{[ 0\beta] }g_{0\alpha}-\hat{R}^g_{(0\beta )}b_{0\alpha}-
\hat{R}^b_{[ 0\alpha] }g_{0\beta}\right) \cr &+ {1\over
g_{00}^2}\left( g_{0\alpha}b_{0\beta}-b_{0\alpha}g_{0\beta}\right) 
\hat{R}^g_{00}\,.\cr}
\ee
Quite recently 
a general proof of the validity of Eqs.~(\ref{ricci1}) has been given 
in Ref.~\cite{haag}, implying that
up to the one loop order Abelian T-duality as defined by Eqs.~(\ref{ndual})
holds, indeed.
Quantum equivalence under duality transformation 
means that the functional integrals computed
with either $T_{ij}^{(0)}(g,b)$ or with $\tilde{T}_{ij}^{(0)}(g,b)$ 
should lead to identical results {\sl to any desired order} in perturbation
theory (for physical quantities of course).
The examples  
studied in Ref.~\cite{bfhp} 
show, however, that in general  
the functional integrals computed with 
$T_{ij}^{(0)}(g,b)$ and $T_{ij}^{(0)}(\tilde g,\tilde b)$ lead to 
different physics beyond one loop.
This also implies that Eqs.~(2.4) of Ref.~\cite{haag} (which are
equivalent to the `naive' duality equations (\ref{ndual}))
will not be consistent for a general background $(g\,,b)$ at the two
loop level.
One might ask, how this (somewhat discouraging) result complies
with the results of Ref.~\cite{rove} showing that duality maps
a conformal field theory (string background) into a string background.
First the examples of Ref.~\cite{bfhp} do not correspond to string
backgrounds, but even more importantly the proof of Ref.~\cite{rove}
is based on a gauging proceedure of chiral currents,
and there is no claim whatsoever that the `naive' transformation
formulae  Eqs.~(\ref{dmetr}) would be exact to all orders in $\alpha'$.
In fact as already mentioned, in 
Ref.~\cite{tse1} for certain string
backgrounds it has been explicitly shown that the Abelian T-duality
transformation rules, Eqs.~(\ref{dmetr}), have to be modified at the two
loop level.

Let us now present our modified transformation rules for the Abelian
T-duality transformations, which should make them a true quantum symmetry,
valid to all orders in perturbation theory. 
Instead of  Eqs.~(\ref{ndual}) we postulate the following equation
for a {\sl finite} mapping, $\gamma(g,b)$:
\be\label{ttilde}
\tilde{T}^{(0)}(g,b)=\left(\gamma^{-1}\circ T^{(0)}\circ\gamma\right)(\tilde{g},
\tilde{b})\,,
\ee
where equality is meant again {\sl modulo 
reparametrizations} of the target space coordinates. If such a
$\gamma$ exists for any $\sigma$-model background ($g$, $b$)
then we would say that the classical duality symmetry is a true
symmetry of the full quantum theory. 
The modified dual (or quantum dual) of a 
$\sigma$-model is defined as
\be\label{cftdual}
(g,b)_q=\gamma(\tilde g,\tilde b)\,.
\ee

Eq.~(\ref{ttilde}) expresses the way in which the renormalization and 
the renormalized 
metric and torsion potential change under a transformation of the bare 
quantities. In this respect Eq.~(\ref{ttilde}) is in complete analogy with the equation
that describes the change in the renormalization of an ordinary {\sl 
parameter},  
whose bare and renormalized values are related as $e_0=Z(e)$. When one
changes from $e_0$ and $e$ to $\hat{e}_0\equiv f(e_0)$ and
$\hat{e}\equiv f(e)$ the relation between the new bare and renormalized 
parameters becomes 
 $\hat{e}_0=f\bigl( Z(f^{-1}(\hat{e}))\bigr)$.
One might be tempted to interpret Eq.~(\ref{ttilde}) as the action of
the duality transformation on the
{\sl renormalized metric} combined with a change of the 
renormalization scheme \cite{RS}. 
It is well known that the $\beta$-functions 
in general are scheme dependent beyond one loop. In our case, however,  
Eq.~(\ref{ttilde}) is only defined for backgrounds possessing an Abelian
isometry, and therefore it is not obvious if the above interpretation
is correct. In fact as shown for a special class of metrics,
Eq.~(\ref{ttilde}) does not correspond to a change of scheme compatible
with {\sl full} target space covariance.

At this point we note that while
it is very natural to assume that
the existence of a non-trivial mapping, $\gamma$, would 
guarantee that duality is 
indeed
a quantum symmetry,
the dual model defined by either sides of
Eq.~(\ref{ttilde}) {\sl does not correspond} any longer to 
a genuine $\sigma$-model as the standard relation between the bare and
the renormalized metric and torsion given by Eqs.~(\ref{fint1}) is lost.
A simple consequence of Eq.~(\ref{ttilde}) for the mapping $\gamma$ is:
\be
\widetilde{\gamma^{-1}}(g)=\gamma(\tilde{g})\,.
\ee
>From Eq.~(\ref{ttilde}) it also immediately follows that the modified duality
transformation maps conformal $\sigma$-models into conformal 
$\sigma$-models in contrast
to the general case. For 
conformal $\sigma$-models the $\beta$-functions vanish, therefore 
the metric and torsion, $(g,b)$, in this case, satisfies
\be\label{cft}
T^{(0)}(g,b)=(g,b)\,,
\ee
where again Eq.~(\ref{cft}) is supposed to hold only modulo a diffeomorphism. 
It is now easy to see that the quantum dual of a 
conformal $\sigma$-model is 
again a conformal $\sigma$-model, indeed. 
By acting with $\gamma$ on both
sides of Eq.~(\ref{ttilde}) one obtains:
\be
\left(\gamma\,\circ\,\tilde{ T}^{(0)}\right)(g,b)=T^{(0)}\left((g,b)_q\right)\,,
\ee
showing that the modified duality transformation
maps conformally invariant models onto themselves.
 
At present we can only analyse Eqs.~(\ref{ttilde}) in the general case
in perturbation theory. This way one determines $\gamma(g,b)$
order by order in the $\alpha^\prime$ expansion, that is
\be\label{gam1}
\gamma_{ij}(g,b)=g_{ij}+b_{ij}+\alpha^\prime M_{ij}(g,b)+\ldots\,.
\ee 
Then Eqs.~(\ref{ttilde}) will connect order by order in $\alpha^\prime$  
$\tilde{T}_{ij}^{(0)}(g,b)$ and $T_{}^{(0)}(\tilde g,\tilde b)$ 
 \footnote{Since in $T_{ij}^{(0)}$
the residues of the higher order poles in $\epsilon$  
 are determined by that
of the single pole, it is enough if Eq.~(\ref{ttilde}) holds order by order
 in 
$\alpha^\prime$ for the residues of single poles on the two sides.}.
Therefore if $M_{ij}$ turns out to be non trivial it means
that the naive duality transformations, Eq.~(\ref{dmetr}), 
must be modified in higher orders.  
Note, 
that $M_{ij}$, the first nontrivial terms of $\gamma_{ij}$,
have no effect on the one loop results: indeed using Eqs.~(\ref{dmetr}),
(\ref{ttilde}) we find
from comparing the coefficients of ${\alpha^\prime/\epsilon}$ 
on the two sides of (\ref{ttilde}) precisely Eqs.~(\ref{ricci1}). In the
next (two loop) order 
the $o({(\alpha^\prime)^2/\epsilon})$ terms in
 Eq.~(\ref{ttilde})
contain both the two loop contributions and the new terms originating from
$M_{ij}$; the right hand side of Eqs.~(\ref{ttilde}) is given as:
\be\label{constr1}
\eqalign{
\hat{Y}_{ij}(\tilde{g},\tilde{b})+&{\delta \hat{R}_{ij}\over\delta g_{kl}}
\vert_{
\tilde{g},\tilde{b}}M_{(kl)}(\tilde{g},\tilde{b})+
{\delta \hat{R}_{ij}\over\delta b_{kl}}\vert_{
\tilde{g},\tilde{b}}M_{[ kl] }(\tilde{g},\tilde{b})\cr
-&{\delta M_{ij}\over\delta g_{kl}}\vert_{
\tilde{g},\tilde{b}}\hat{R}_{(kl)}(\tilde{g},\tilde{b})-
{\delta M_{ij}\over\delta b_{kl}}\vert_{
\tilde{g},\tilde{b}}\hat{R}_{[ kl] }(\tilde{g},\tilde{b})\,.\cr}
\ee
Equating these with the $o({(\alpha^\prime)^2/\epsilon})$ terms on the
left hand side of (\ref{ttilde}) -- which can be obtained from the
expressions on the
r.h.s. of Eq.~(\ref{ricci1}) by replacing the various components of
 $\hat{R}_{ij}(g,b)$
by the corresponding components of $\hat{Y}_{ij}(g,b)$ --  
yields an equation for $M_{ij}(g,b)$.
We leave the analysis of the resulting equations (the problem of existence
of a solution for a general background) for future work
as it is somewhat complicated. We shall content ourselves 
to present below the solution just for the special case of a block diagonal
metric 
which is fairly simple from a calculational point of view, but
shows that our modified duality equations (\ref{ttilde}) admit a
non trivial solution.

We would like to emphasize once again that 
the existence of a non-trivial $M_{ij}$ ( necessary for duality be 
a true quantum symmetry), implies that
the dual model cannot be interpreted as
a standard $\sigma$-model {\sl beyond one loop}, hence the modified
duality transformation {\sl does not map} $\sigma$-models into
$\sigma$-models.

Let us now present an explicit construction of the mapping $\gamma$
for a special class of $\sigma$-models, which are
the  `block diagonal' purely metric $\sigma$-models.
In these models  
$b_{ij}$ and consequently the antisymmetric part of $T_{ij}^{(0)}$ 
vanishes identically and only the $g_{00}$ and the $g_{\gamma\delta}$
 components 
of the metric (and 
of $T_{ij}^{(0)}$ ) are different from zero in the adapted coordinate system: 
$g_{00}=
g_{00}(\xi^\gamma)$, $g_{0\gamma}\equiv 0$ and 
$g_{\gamma\delta}=g_{\gamma\delta}(\xi^\beta)$. For these models 
the equations following from the modified duality transformation rules
Eqs.~(\ref{ttilde})
for $M_{ij}$ simplify considerably:
\be\label{constr2}
\eqalign{
-{\hat{Y}_{00}(g)\over g_{00}^2}&=\hat{Y}_{00}(\tilde{g})
+{\delta R_{00}\over\delta g_{kl}}\vert_{
\tilde{g}}M_{kl}(\tilde{g})
-{\delta M_{00}\over\delta g_{kl}}\vert_{
\tilde{g}} R_{kl}(\tilde{g})+\eta^\alpha\pa_\alpha\tilde{g}_{00}\,,\cr
\hat{Y}_{\alpha\beta}(g)&=\hat{Y}_{\alpha\beta}(\tilde{g})
+{\delta R_{\alpha\beta}\over\delta g_{kl}}\vert_{
\tilde{g}}M_{kl}(\tilde{g})
-{\delta M_{\alpha\beta}\over\delta g_{kl}}\vert_{
\tilde{g}} R_{kl}(\tilde{g})+D_\alpha\eta_\beta+D_\beta\eta_\alpha\,,
\cr}
\ee
where $R_{ij}$ denote the ordinary Ricci tensor and 
$\eta^\alpha$ describe the above mentioned reparametrization 
(diffeomorphism) freedom. Eqs.~(\ref{constr2}) admit the following
simple solution:
\be\label{sol1}
M_{00}(g)={g_{00}\over2}(\pa_\alpha\ln(g_{00}) )^2\,,\qquad M_{0\beta}=0\,,
\qquad  
M_{\beta\alpha}=0\,,
\ee
where $(\pa_\beta\ln g_{00})^2$ stands for $g^{\beta\alpha}
\pa_\beta\ln g_{00}\pa_\alpha\ln g_{00}$ and
$\eta_\alpha=\pa_\alpha (\pa_\beta\ln g_{00})^2/8$.
This clearly shows that our proposed modication of the Abelian T-duality
transformations (\ref{ttilde}) is nontrivial. Eqs.~(\ref{sol1}) coincides with 
the two loop modification found in Ref.~\cite{tse1}. 

Knowing an explicit solution of Eq.~(\ref{ttilde}) 
it is not 
difficult to see that it does not
corresponds to a simple change of the renormalization scheme 
in the sense of Ref.~\cite{RS}.
In fact there is no choice of the constants $k_1$ and $k_2$ in 
\be
\hat{g}_{ij}=g_{ij}+\alpha^\prime (k_1R_{ij}+k_2g_{ij}R),
\ee
describing the most general change in the renormalization
scheme compatible with {\sl full target space covariance}
that would reduce to Eqs.~(\ref{sol1}) (even up to reparametrizations) 
in our block diagonal special case.

\section{Examples}

The quantum equivalence of dually
related models can be studied on the simple but not completely trivial 
example of the $O(3)$ $\sigma$-model described in terms of polar coordinates:
\be\label{freelag1}
{\cal L}={1\over2\lambda}\left((\pa_\mu \theta)^2+
\sin^2\theta (\pa_\mu\phi)^2\right)={1\over\lambda}\tilde{\cal L},
\ee
and its abelian dual (based on the $\phi$ translation isometry):
\be\label{freelag2}
{\cal L}^d={1\over2\lambda}\left((\pa_\mu \theta)^2
+{(\pa_\mu\Phi)^2\over\sin^{2}(\theta)}\right)
={1\over\lambda}\tilde{\cal L}^d\,.
\ee
This model is already sufficient to illustrate
some of the main points of this paper. Eq.~(\ref{freelag1}) describes an 
asymptotitcally free model and we demonstrate below that one 
indeed has to use
the modifications of the duality transformations, Eqs.~(9,16,17), to  
obtain the same $\beta$-function from the dual model as from the original one.
    
We now
carry out explicitly the renormalization up to two loops of both the original
(\ref{freelag1}) and the dual theory (\ref{freelag2}) to see if
$\lambda$ really gets renormalized in the same way. 
Our general strategy to carry out the renormalization of 
this type of $\sigma$-models
and to obtain the corresponding $\beta$ functions is described 
quite in some detail
in Ref.~\cite{bfhp}, here we just quote the essential formulae.
The procedure
is based on the one resp.~two
loop counterterms for the general $\sigma$-models
Eqs.~(\ref{fint1},\ref{y}).
The 
loop expansion parameter, $\alpha^\prime$, expressed in terms of
the coupling, $\lambda$, is $\alpha^\prime=\lambda/(2\pi)$.
The explicit form of the one and two loop counterterms can then be written as:
\be\label{fctms1}
\Sigma_1={1\over4}\hat{R}_{ij}\Xi^{ij},\qquad
\Sigma_2={1\over8}\hat{Y}_{ij}\Xi^{ij}.
\ee
We convert the
previous counterterms into coupling renormalization
by assuming that in the one ($i=1$) and two ($i=2$) loop orders
their bare and renormalized values are related as
\be\label{couplren}
\lambda_0=\mu^\epsilon\lambda\Bigl( 1+{\zeta_1\lambda\over\pi\epsilon}
+{\zeta_2\lambda^2\over8\pi^2\epsilon}+...\Bigr)
=\mu^\epsilon\lambda Z_{\lambda}(\lambda)\,,
\ee
where the dots stand for both the higher loop contributions and for the
higher order pole terms.
The unknown $\zeta_i$ ($i=1$,$2$)
are determined from the following equations:
\be\label{renormeqs}
-\zeta_i\tilde{\cal L}+
{\delta \tilde{\cal L}\over\delta\xi^k}\xi^k_i(\xi)=\Sigma_i\,.
\ee
As discussed in Ref.~\cite{bfhp}
Eqs.~(\ref{renormeqs}) admits a simple
interpretation: the general counterterms 
may be absorbed by the renormalization of the
coupling together with
a (in general non-linear) redefinition of the fields $\xi^j$:
\be\label{renormxi}
\xi^j_0=\xi^j+{\xi^j_1(\xi^k)\lambda\over\pi\epsilon}
+{\xi^j_2(\xi^k)\lambda^2\over8\pi^2\epsilon}+...\;,
\ee
where
$\xi^j_1$, $\xi^j_2$ have to satisfy Eqs.~(\ref{renormeqs}).
In the special case when $\xi_i^k$ depends
linearly on $\xi $ i.e.\  $\xi_i^k(\xi)=\xi^k y_i^k$,
Eqs.~(\ref{renormxi})
simplify to an ordinary multiplicative wave function renormalization.
We emphasize that it is not a priori guaranteed that Eqs.~(\ref{renormeqs})
may be solved at all
for $\zeta_i$ and the functions $\xi^k_i(\xi)$.
If Eqs.~(\ref{renormeqs}) do not have a solution, then
the renormalization of the model is not possible within the
restricted subspace
characterized by the coupling $\lambda$ 
in the (infinite dimensional)
space of  metrics.
On the other hand, if Eqs.~(\ref{renormeqs}) admit a solution, then, writing 
$Z_\lambda=1+y_\lambda (\lambda)/\epsilon+...$  
the $\beta$ function of $\lambda$ is readily obtained:
\be\label{betadef}
\mu{d\lambda\over d\mu}=
\beta_\lambda=\lambda^2{\pa y_\lambda\over\pa\lambda}\,. 
\ee
For the $O(3)$ $\sigma$-model the explicit form of the counterterms 
\be
\Sigma_1={1\over4}\left((\pa_\mu \theta)^2+
\sin^2\theta(\pa_\mu\phi)^2\right),\qquad \Sigma_2=2\Sigma_1, 
\ee
implies, that the $\theta$ and $\phi$ fields undergo no renormalization. 
Eqs.~(\ref{renormeqs}) give in this case $\zeta_1=-{1\over2}$, $\zeta_2=-1$,
and using them in Eq.~(\ref{betadef}) leads to the well known $\beta$-function
: $\beta_\lambda=-{\lambda^2\over2\pi}(1+{\lambda\over2\pi})$.   

For the dual model, Eq.~(\ref{freelag2}), the counterterms have a slightly 
more complicated form:
\be\label{ddualcont}
\eqalign{
\Sigma_1&=-{1\over4}{1+\cos^2\theta\over\sin^2\theta}\left((\pa_\mu \theta)^2+
{(\pa_\mu\Phi)^2\over\sin^2\theta}\right),\cr
\Sigma_2&={1\over2}{(1+\cos^2\theta)^2\over\sin^4\theta}\left((\pa_\mu \theta)^2+
{(\pa_\mu\Phi)^2\over\sin^2\theta}\right),\cr} 
\ee
Looking at these counterterms we see that in principle 
in the present renormalization problem we can have an 
ordinary wavefunction renormalization for $\Phi$ 
and a redefinition of the variable $\theta$, i.e. we have:  
\be\label{fctms2}
\eqalign{
\lambda_0=&\mu^{\epsilon }\lambda\Bigl(1+
{\zeta_1\lambda\over\pi\epsilon}+
{\zeta_2\lambda^2\over8\pi^2\epsilon}+..\Bigr),\cr 
\theta_0=&\theta+{T_1(\theta)\lambda\over\pi\epsilon}+
{T_2(\theta)\lambda^2\over8\pi^2\epsilon}+...\cr
\Phi_0=&\Phi\Bigl(1+
{z_1\lambda\over\pi\epsilon}+
{z_2\lambda^2\over8\pi^2\epsilon}+..\Bigr).\cr}
\ee
At one loop Eqs.~(\ref{renormeqs}) yields now the following equations:
\be\label{f1loop}
2T_1^\prime-\zeta_1=-{1+\cos^2\theta\over2\sin^2\theta},\qquad
{2z_1-\zeta_1\over\sin^2\theta}-{2T_1\cos\theta\over\sin^3\theta}=
-{1+\cos^2\theta\over2\sin^4\theta}\,,   
\ee
while at two loops one obtains:
\be\label{f2loop}
2T_2^\prime-\zeta_2={(1+\cos^2\theta)^2\over\sin^4\theta},\qquad
{2z_2-\zeta_2\over\sin^2\theta}-{2T_2\cos\theta\over\sin^3\theta}=
{(1+\cos^2\theta)^2\over\sin^6\theta}\,.
\ee
The two equations appearing in (\ref{f1loop}) have a consistent solution
\be 
T_1(\theta)={\rm cotg}\theta /2\,,\qquad \zeta_1=-1/2\,,\qquad z_1=-1/2\,,
\ee
which shows that at one loop the $\beta$ functions of (\ref{freelag1})
and of (\ref{freelag2}) are indeed the same,  
while at the two loop order we meet the problem exhibited already
in several examples in Ref.~\cite{bfhp}, that is
there is no choice of $\zeta_2$ and $z_2$ that would guarantee that $T_2$
expressed algebraically from the second equation in (\ref{f2loop}) would  
also solve the differential equation in (\ref{f2loop}). 
Thus renormalizing the dual model 
described by Eq.~(\ref{freelag2}) as a standard $\sigma$-model one
finds that it 
is not renormalizable, therefore
it cannot be 
equivalent to the $O(3)$ model given by Eq.~(\ref{freelag1}).

The $O(3)$ $\sigma$-model belongs to the class of block diagonal
purely metric $\sigma$-models therefore we now apply the modified
duality transformation Eqs.~(\ref{sol1}) to demonstrate explicitly
how the two loop `anomaly' is removed in our framework.
In fact 
taking into account the explicit modification of Eqs.~(\ref{dmetr}) following
from Eqs.~(\ref{ttilde}, \ref{sol1})
 changes the two loop equations, Eqs.~(\ref{f2loop}), as:
\be\label{f2loopm}
2T_2^\prime-\zeta_2=1-{4(1+2\cos^2\theta)\over\sin^4\theta},\quad
{2z_2-\zeta_2\over\sin^2\theta}-{2T_2\cos\theta\over\sin^3\theta}=
-{\sin^4\theta+4\cos^2\theta\over\sin^6\theta}\,.
\ee
Remarkably this system admits a consistent solution:
\be
T_2(\theta)=2{\cos\theta\over\sin^3\theta}\,, \qquad\zeta_2=-1\,,\qquad 
z_2=-1,
\ee
showing that in the new framework 
 the dual of the $O(3)$ $\sigma$-model leads to the same 
$\beta$-function as the original model 
even at 
the two loop level, and that the modifications of the `naive' Abelian
duality transformation rules, Eqs.~(\ref{dmetr}), are essential, indeed. 

Finally we show that as far as the coupling constant renormalization is 
concerned, the non trivial $\gamma$ mapping, Eq.~(\ref{sol1}), 
is also necessary to establish the two loop equivalence between 
\be\label{frlag1}
{\cal L}={1\over2\lambda}\left((\pa_\mu r)^2+r^2(\pa_\mu\phi)^2\right)
\ee
describing two free scalar fields in polar coordinates,  
and its abelian dual:
\be\label{frlag2}
{\cal L}^d={1\over2\lambda}\left((\pa_\mu r)^2+r^{-2}(\pa_\mu\Phi)^2\right)
={1\over\lambda}\tilde{\cal L}^d\,.
\ee
Eq.~(\ref{frlag1}) describes a (not very complicated) CFT and some
evidence was given in Ref.~\cite{kiri}, using the \lq\lq minisuperspace" 
approximation, that the dual model is indeed 
equivalent to the original one (see also Ref.~\cite{giveon}). 
In our framework the equivalence hinges upon
whether the coupling constant of the dual theory really gets renormalized
as indicated by the non trivial metric or stays unrenormalized as in the 
original free model. In applying the coupling constant renormalization  
we note that 
Eq.~(\ref{frlag2}) is invariant under the $\lambda\rightarrow a\lambda$,
$r\rightarrow a^{1/2}r$, $\Phi\rightarrow a\Phi$ scaling transformations, 
thus we can effectively set the wavefunction renormalization of $\Phi$ to 
one, i.e. we have:  
\be\label{nfctms2}
\eqalign{
\lambda_0=&\mu^{\epsilon }\lambda\Bigl(1+
{\zeta_1\lambda\over\pi\epsilon}+
{\zeta_2\lambda^2\over8\pi^2\epsilon}+..\Bigr),\cr 
r_0=&r+{r_1(r)\lambda\over\pi\epsilon}+
{r_2(r)\lambda^2\over8\pi^2\epsilon}+...\cr}
\ee
At one loop we get from Eqs.~(\ref{renormeqs}) a system that admits the 
solution $r_1(r)=1/(2r)$, $\zeta_1=0$, however at two loops it yields
\be\label{nf2loop}
2r_2^\prime-\zeta_2={4\over r^4},\qquad\qquad 
-{\zeta_2\over r^2}-{2r_2(r)\over r^3}=
{4\over r^6}\,,
\ee
which admits {\sl no} solution. Taking into account the terms coming from the 
non trivial part of the $\gamma$ mapping changes Eqs.~(\ref{nf2loop}) to
\be\label{nf2loopm}
2r_2^\prime-\zeta_2={9\over r^4}\,,\qquad\qquad 
-{\zeta_2\over r^2}-{2r_2(r)\over r^3}=
{3\over r^6}\,,
\ee 
which admits a consistent solution:
$r_2(r)=-3/(2r^3)$, $\zeta_2=0$; 
showing that the dual of the free model remains `free' even at 
the two loop level. 

Based on the above (admittedly as yet incomplete) evidence, that
our proposed modified duality transformations, (\ref{ttilde}),
do restore the equivalence between dual $\sigma$-models
in perturbation theory, we expect that for the general case
(i.e.\ a not necessarily block diagonal metric tensor $g_{0\alpha}\ne 0$ 
and $b_{ij}\ne 0$) Eqs.~(\ref{ttilde}) also admit a solution.
Then it is natural to conjecture, that the  
modified duality transformations restore the equivalence between the 
original and the dual models at two loops for the
example discussed in Sect.~4.2 of Ref.~\cite{bfhp} just as in the $O(3)$ case.
Furthermore we also expect that a similar modification of 
the non-Abelian duality transformations restore the two-loop
equivalence between the principal $\sigma$-model and its nonabelian
dual (for the two-loop problem in that case, see Sect.~5 of Ref.~\cite{bfhp}).
(Recently the quantum equivalence between gauged WZW models
and their non-abelian duals has been proved in Ref.~\cite{hew}.)
Finally we think it would be interesting to investigate 
the $\gamma$ mapping
beyond two loops.

\end{document}